\documentstyle[prl,aps]{revtex}
\twocolumn
\begin{document}
\input{epsf}
\draft
\twocolumn[\hsize\textwidth\columnwidth\hsize\csname@twocolumnfalse\endcsname
\title{Area-preserving dynamics
of a long slender finger by curvature: a test case for the globally
conserved phase ordering}
\author{Avner Peleg, Baruch Meerson, and Arkady Vilenkin}
\address{Racah Institute  of  Physics, Hebrew
University   of  Jerusalem, Jerusalem 91904, Israel}
\author{Massimo Conti}
\address{Dipartimento di Matematica e Fisica, Universit\'{a} di
Camerino, and Istituto Nazionale di Fisica della Materia, 62032,
Camerino, Italy}
\maketitle
\begin{abstract}
A long and slender finger can serve as a
simple ``test bed''
for different phase ordering models. In this work,
the globally-conserved,
interface-controlled dynamics of a long finger  is investigated,
analytically and numerically,
in two dimensions. An important limit is considered when
the finger dynamics are reducible to the area-preserving motion by
curvature. A free
boundary problem for the finger shape is formulated.
An asymptotic perturbation
theory is developed
that uses the finger aspect ratio as a small
parameter. The leading-order
approximation is a modification of
``the Mullins finger" (a well-known analytic solution)
which width is allowed to slowly vary with time. This
time
dependence
is described, in the leading order, by an exponential law with the
characteristic time proportional to the (constant) finger area.
The subleading terms of the asymptotic theory
are also calculated. Finally, the finger
dynamics  is investigated numerically, employing
the
Ginzburg-Landau equation with a global conservation law. The
theory is in a very good agreement with the numerical solution.
\end{abstract}
 \vskip1pc]
\narrowtext

\section{INTRODUCTION}
This work is motivated by the recent developments in
phase-ordering theory. Phase ordering is the growth of order from
disorder via domain growth and coarsening. As phase ordering
systems are strongly nonlinear and highly disordered, their
theoretical description remains  challenging \cite{Bray}.  To  get
insight, it is often useful to consider the coarsening dynamics of
simple test objects. One such object is a spherical droplet of the
``minority'' phase. It serves as the building block of simplified
phase-ordering theories for locally conserved
(bulk-diffusion-controlled) \cite{LS} and globally conserved
(interface-controlled) \cite{Wagner} systems.

Being too a simple object, a spherical droplet may not give
sufficient insight. For example, it's dynamics (shape-preserving
shrinking or expansion) look the same in both locally, and
globally conserved systems. On the other hand, there are important
differences in the phase-ordering behavior of locally and globally
conserved systems. In both cases, an ensemble of droplets exhibits
Ostwald ripening, and the corresponding
mean-field theories, due to Lifshitz and Slyozov
\cite{LS} and Wagner \cite{Wagner}, respectively, predict dynamic scaling
behavior of the droplet size distribution
(with different dynamic exponents).  However, for
finite volume fractions, the difference in the type of conservation
law leads to a different role of correlations.
Correlations between neighboring
droplets are much more important in the locally-conserved systems
\cite{Marder}, than in the globally-conserved
ones \cite{MS}. Even bigger differences have been found in the
phase ordering dynamics of locally and globally conserved
systems with long-range correlations.
Coarsening of {\it fractal clusters} shows dynamic scale
invariance and ``normal'' scaling in the case of global conservation
\cite{Peleg1}, and breakdown of scale invariance and anomalous
scaling in locally conserved systems \cite{CMS1,CMS2}.

It is natural to attribute
these differences to a basic difference in the character of transport:
in
globally
conserved systems the transport is uninhibited by Laplacian screening
effects,
typical for locally conserved diffusion-controlled
systems. To study this basic difference
in a simple setting,
the dynamics of a single long and slender
finger is often considered. It has been found
recently \cite{CMS2} that, in the
locally-conserved, diffusion-controlled system, the finger acquires
a dumbbell shape and shows non-trivial dynamic
scalings for the finger length
and the ``ball'' size, while the initial finger width remains
almost constant until a late stage of the dynamics. Furthermore, a
finger-shaped domain (a ``needle'')
served
as a test object in still another conserved
coarsening system: the one controlled by
edge diffusion
\cite{Olivi-Tran}. Finger-like objects appear
naturally in two-dimensional simulations of dewetting \cite{Koplik}, etc.
In this work we
investigate the globally conserved dynamics of a long slender finger.
An additional
motivation for studying the finger dynamics is a recent observation that,
at a late stage of coarsening of fractal clusters, the cluster
morphology shows long branches, or fingers,
in both locally \cite{CMS1,CMS2,Irisawa1}, and
globally \cite{Peleg1,Irisawa2} conserved systems.

Here is an outline of the rest of the paper. In Section 2 we
briefly review the Ginzburg-Landau equation with a global
conservation law (a phase field model for globally-conserved phase
ordering) and its general sharp interface formulation in two
dimensions. Under a certain condition (that will be elucidated)
this formulation is reduced to the area-preserving dynamics by
curvature. In Section 3 we formulate a moving boundary problem for
the finger dynamics and develop a perturbation theory which
enables us to obtain an intermediate asymptotic solution for the
finger shape. In Section 4 we return to the globally-conserved
Ginzburg-Landau equation and investigate the finger dynamics
numerically. Section 5 includes a brief discussion.

\section{globally conserved phase ordering: a phase field model and
its sharp-interface limit}

Globally-conserved phase ordering can be described by a simple phase
field
model \cite{Rubinstein,Majumdar,Rutenberg}. This model represents the
dynamics in terms of a simple gradient descent
\begin{equation}
\frac{\partial u}{\partial t} = - \frac{\delta F}{\delta u}
\label{descent}
\end{equation}
with the Ginzburg-Landau free
energy functional
\begin{equation}
F[u] = \int \left[(1/2) (\nabla u)^2 + V(u) + H u \right] d {\bf
  r}
\label{F}
\end{equation}
and a double-well potential $V(u) = (1/4) (1-u^2)^2$. The
effective ``magnetic field'' $H=H(t)$ varies in time so as to
impose global conservation law
\begin{equation}
<u({\mathbf{r}},t)>=\frac{1}{L_{x}L_{y}}\int u({\mathbf{r}},t)\,
d\,{\mathbf{r}}=\mbox{const}
 \,.
\label{51}
\end{equation}
Here $u({\mathbf{r}},t)$ is the order parameter field, $L_{x}\gg 1$ and
$L_{y}\gg 1$ are the linear dimensions of the system (a
two-dimensional
rectangular
box), and the integration is carried out over the entire box.
Eqs. (\ref{descent})-(\ref{51})
yield the nonlocal Ginzburg-Landau equation
\begin{equation}
\frac{\partial u}{\partial t}=\nabla^{2} u+u-u^{3}- <u-u^3> \,.
\label{GL}
\end{equation}
Either no-flux, or periodic boundary
conditions are assumed.

For a wide class of initial conditions this coarsening
system segregates, at late
times, into large
domains of "phase 1" and "phase 2" separated by thin domain walls (which
width is
of order unity) \cite{MS,Rubinstein,Mikhailov}.
Correspondingly, a sharp interface theory  can be
developed for these late times \cite{MS}. At these times
the magnetic field
$H(t)$ is already small,
$H(t)\ll 1$, and slowly varies with time. The phase
field in the phases $1$ and $2$ is almost uniform and rapidly
adjusts to the
current value of $H(t)$, so that $u=-1-H(t)/2$
and $1-H(t)/2$, respectively. The normal velocity of the
interface is given by \cite{MS}
\begin{equation}
v_{n}(s,t)=\kappa(s,t)-\frac{3}{\sqrt{2}} H(t)
 \,,
\label{52}
\end{equation}
where $\kappa$ is the local curvature and $s$ is the coordinate
along the interface. The positive sign of $v_n$ corresponds to the
interface moving towards phase 1, while $\kappa$ is positive when
the interface is convex toward phase 2.

The dynamics of $H(t)$ is determined
by \cite{MS}
\begin{equation}
\dot{H}(t)=\frac{4\Lambda(t)}{L_{x}L_{y}}
\left(\overline{\kappa(s,t)}-\frac{3}{\sqrt{2}} H(t)\right)
 \,,
\label{53}
\end{equation}
where
\begin{equation}
\overline{\kappa(s,t)}=\frac{1}{\Lambda(t)}\oint \kappa(s,t)\,ds
 \,,
\label{average}
\end{equation}
and $\Lambda(t)$ is the total interface length. Equations
(\ref {52}) and (\ref {53}) make a complete set and
provide a general sharp-interface formulation to this problem.
In some cases
this formulation can be further simplified \cite{MS,Rubinstein}.
Let us compute the rate of change of the domain area, $\dot{A}$:
\begin{equation}
\dot{A}(t) = \oint v_{n}(s,t)\,ds = \Lambda(t)
\left[\overline{\kappa(s,t)}-\frac{3}{\sqrt{2}}H(t)\right] \,.
\label{12}
\end{equation}
If the two terms in the right hand side of Eq. (\ref{12})
balance
each other,
\begin{equation}
H(t) \simeq \frac{\sqrt{2}}{3}\, \overline{\kappa(s,t)}\,,
\label{H}
\end{equation}
the domain area remains approximately constant. Then, using
Eqs. (\ref{52}) and (\ref{H}), we obtain
\begin{equation}
v_{n}(s,t)=\kappa(s,t)-\overline{\kappa(s,t)} \,. \label{10}
\end{equation}
Model (\ref{10}) is known as area-preserving flow by curvature
\cite{Gage}. Dynamics (\ref{10}) reduce the interface length of
the system \cite{Rubinstein,Gage}, leaving the areas of each of
the phases constant. The only stable steady state of Eq.
(\ref{10}) (not attached to the boundaries of the system) is a
single perfectly circular domain \cite{Rubinstein}. Under what
condition is model (\ref{10}) a good approximation to the more
general sharp-interface model (\ref{52}) and (\ref{53})? Consider
Eq. (\ref{53}) which includes the same combination
$\overline{\kappa}-(3/\sqrt{2}) H(t)$. Let us treat the term in
the right hand side of Eq. (\ref{53}) perturbatively, putting
$H(t)=H^{(0)}(t)+h(t)$, where $H^{(0)}(t)=(\sqrt{2}/3)\,
\overline{\kappa}$ is the leading term, and $h(t)$ is a subleading
term. Keeping terms up to order $h(t)$ we obtain:
\begin{equation}
h(t)= - \frac{\sqrt{2}L_{x}L_{y}}{12\Lambda}\,\dot{H}^{(0)}(t)=
-\frac{L_x L_y}{18 \Lambda}\, \dot{\overline{\kappa}} \label{54a}
\end{equation}
If there is only one (single-connected) domain, $\overline{\kappa}
=-2\pi/\Lambda$. Therefore,
\begin{equation}
h(t)=
-\frac{\pi
L_{x}L_{y}}{9}\frac{\dot{\Lambda}}{\Lambda^{3}}
 \,.
\label{55}
\end{equation}
Requiring $|h(t)|\ll |H^{(0)}(t)|$, we obtain the ``area-preservation''
criterion
\begin{equation}
\frac{L_{x}L_{y}|\dot{\Lambda}|}{\Lambda^{2}}\ll 1
 \,.
\label{56}
\end{equation}
In Section 3 we will investigate the area-preserving dynamics
of a long slender finger, and verify criterion (\ref{56})
{\it a posteriori}.

\section{Area-preserving finger dynamics: a perturbation theory}
Let the initial condition for dynamics (\ref{10})
represent a long and narrow rectangular bar
with length $2a_{0}$ and width $2\Delta_{0}$. We will see
that, with time, the bar evolves into a finger-shaped object.
We place the origin
of a Cartesian coordinate system in the finger's center and denote
by $x$ the coordinate along the finger and by $y=y(x,t)$ the
instantaneous location of the finger boundary. Because of the
symmetry with respect to the $x$-axis, we will be interested only
in the upper boundary dynamics: $y(x,t)\geq 0$.
In terms of $y(x,t)$ the local curvature is
given by \cite{Courant}:
$$
\kappa(s,t)=\frac{y_{xx}}{(1+y_{x}^{2})^{3/2}}\,,
$$ where $y_x
\equiv \partial y(x,t)/\partial x$. The average curvature
$\overline{\kappa(s,t)}$ is
$$\overline{\kappa(s,t)}= \frac{-2\pi}{\Lambda}=
\frac{-\pi}{2\int_{0}^{a(t)}(1+y_{x}^2)^{1/2} dx}\,,$$ where
$a(t)$ is the time-dependent position of the finger tip.
Since
$$
\frac{\partial y}{\partial t}=(1+y_{x}^{2})^{1/2}v_{n}\,,
$$
dynamics (\ref {10}) imply
the following evolution equation for $y(x,t)$:
\begin{equation}
\frac{\partial y}{\partial t} =\frac{y_{xx}}{1+y_{x}^{2}}+ \frac{
\pi\,(1+y_{x}^2)^{1/2}}{2\int_{0}^{a(t)}(1+y_{x}^2)^{1/2} dx} \,.
\label{11}
\end{equation}
$a(t)$ is defined by the
boundary conditions $y (x=a,t)=0$ and $y_x (x=a,t) = - \infty$. An
additional boundary condition follows from the symmetry with
respect to the $y$-axis: $y_x (x=0,t)=0$.  Equation (\ref{11}), a
nonlinear integro-differential equation in partial derivatives,
describes a free-boundary problem. It is easy to check that the
dynamics (\ref{11}) preserve the finger area $A$.

It can be proved  \cite{Rubinstein}
that any convex domain evolving by Eq. (\ref{10}) [and therefore by
Eq. (\ref{11})] finally becomes a circle
with the same area $A$. We are interested in
the finger dynamics at times much smaller than those
needed for approaching
this equilibrium (but still large enough so that
details of the initial
shape are forgotten). The corresponding strong double inequality will
be presented at the end of this Section.

As long as the ratio of the finger width to its length is small,
we can use it as the small parameter of our theory. Indeed, near the
finger tip
the
first (local) term in the right hand side of Eq. (\ref{11}) is of
the order of the inverse finger width, while the second (nonlocal)
term is of the order of the inverse finger length. Therefore, the
nonlocal term can be treated perturbatively. The unperturbed
equation,
\begin{equation}
\frac{\partial y}{\partial t} =\frac{y_{xx}}{1+y_{x}^{2}}
 \,,
\label{13}
\end{equation}
was obtained more than 40 years ago by Mullins in the context of
motion of grain boundaries \cite{Mullins}. It corresponds to a
{\it non-conserved} motion by curvature: $v_{n}(s,t)=\kappa$. We
notice immediately that the perturbed problem, Eq. (\ref{11}),
possesses an integral of motion (the finger area) which is absent
in the corresponding unperturbed problem. This situation is
unusual: more often than not perturbations destroy integrals of motion
pertinent to the unperturbed system. This unusual property will be
exploited in the following, as we will require area conservation
in each order of the perturbation expansion.

Mullins \cite{Mullins} obtained a one-parameter family of
traveling-wave solutions of Eq. (\ref{13}) describing a
half-infinite constant-width finger retreating by its tip's
curvature:
\begin{equation}
y=Y(x,t)=\frac{1}{c}\arccos\,\left[\exp\left(-c\,\xi \right)\right]
 \,,
\label{14}
\end{equation}
where $\xi=a(t)-x$, $a (t) = a_0  - c t$ is the time-dependent
position of the finger tip, and the constant speed of retreat $c$ is
the parameter of this family of solutions.
The finger's half-width $\Delta=\pi/(2\,c)$ is constant. We will show
that
this solution
(with two important modifications) can serve as the leading (or
zero) order
approximation of our perturbation theory for the area-preserving finger
dynamics. First, we introduce a (slow) time-dependence in the
parameter $c$. Second, we ``stick together''
two identical (very long but finite)
half-fingers. The
resulting ansatz is the following:
\begin{equation}
Y(x,t) = \frac{1}{c(t)}\arccos\,\left[\exp\left(-c(t)\,\xi
\right)\right] \,,
 \label{14a}
\end{equation}
where $\xi =a(t)-|x|$, and $a(t)>0$ is yet unknown. The
finger half-width $\Delta (t) = \pi/2 c(t)$ is now time-dependent;
its growth with time makes up, in the area conservation, for the
finger shortening. As a function of $x$, ansatz  (\ref{14a})
is continuous everywhere. Its $x$-derivative is not
continuous at $x=0$. However, as far as $\epsilon (t) = \Delta
(t)/a (t)$ is very small: $\epsilon(t) \ll 1$, the $x$-derivatives
at $x=0$ from left and right are of order of $\exp (-
1/\epsilon)$, that is exponentially small. This and other
exponentially small effects will be neglected throughout the
paper.
Because of the symmetry with respect to the $y$-axis, we will
consider only the region $0\leq x\leq a(t)$. The perturbation
expansion for $y(x,t)$ is
\begin{equation}
y(x,t)=Y(x,t)+\delta y(x,t)+ \dots
 \,,
\label{16}
\end{equation}
where $\delta y(x,t)$ is the subleading (or first order) term. Inserting
Eq. (\ref {14a}) into Eq. (\ref{11})
and keeping only the zero order terms, we obtain
\begin{equation}
\dot{a}=-c=-\pi/(2\,\Delta)\,. \label{adot}
\end{equation}
The evolution equation for $\Delta(t)$ follows from the requirement
of area conservation in the zero order. Calculating the area under
the $Y(x,t)$ profile, we obtain:
\begin{equation}
\int_{0}^{a(t)}Y(x,t)\,dx=a\Delta-\frac{2\Delta^{2}\ln
2}{\pi}+O(\exp\,(-1/\epsilon))
 \,.
\label{16a}
\end{equation}
One can see that the finger area in the zero order is simply
$A=4\,a\,\Delta$. Demanding $\dot{A}=0$ and using Eq.
(\ref{adot}), we obtain:
\begin{equation}
\dot{\Delta}=\pi/(2\,a)\,. \label{Ddot}
\end{equation}
We will see later that the same equation follows from
the analysis of the first order correction. The solution of the
zero order equations (\ref{adot}) and (\ref{Ddot}) is
\begin{equation}
a^{(0)}(t)=a_{0}\exp\,(-\pi t/2a_{0}\Delta_{0})
 \,,
\label{40a}
\end{equation}
and
\begin{equation}
\Delta^{(0)}(t)=\Delta_{0}\exp\,(\pi t/2a_{0}\Delta_{0})
 \,.
\label{40b}
\end{equation}
Therefore, in the zero order, the
finger shortens (and its
width grows) exponentially with time. The characteristic growth time
is of order of the finger area. The zero order solution
yields a criterion for the validity of the perturbation theory.
The aspect ratio of the evolving
finger should be very small which leads to
\begin{equation}
t \ll a_{0}\Delta_{0}\ln\frac{a_{0}}{\Delta_{0}}
 \,,
\label{41}
\end{equation}
This condition sets an upper limit for the times for which our
perturbation expansion is valid.

We now turn to calculating the small correction $\delta y(x,t)$.
Since the nonlocal term of Eq. (\ref {11}) is already of order
$\epsilon$, we evaluate it on the zero-order solution $Y(x,t)$
which yields $\pi\, (2a)^{-1}(1+Y_{x}^{2})^{1/2}$. Introduce a new
variable $u$:
\begin{equation}
u(x,t)=\exp\,(-c(t)\xi) \,. \label{18}
\end{equation}
Substituting Eq. (\ref {16}) into Eq. (\ref{11}) and linearizing,
we get a linear partial differential equation for $\delta y(x,t)$:
\begin{eqnarray}
\delta y_{t} & = &(1-u^{2})\,\delta y_{xx}-2\,c\,u^{2}\,\delta
y_{x}\nonumber\\ & & -\frac{(c+\dot{a})\,u}{(1-u^{2})^{1/2}}+
\frac{\dot{c}}{c^{2}}\,q(u)+\frac{\pi}{2\,a\,(1-u^{2})^{1/2}}\,,
\label{19}
\end{eqnarray}
where
\begin{equation}
q(u)=\frac{u\ln\,u}{(1-u^{2})^{1/2}}+\arccos\,u \,.
 \label{19a}
\end{equation}
It is convenient to go over to new independent variables $\xi$ and
$t$. We define $\delta y(x,t)=f(\xi,t)$ and obtain $\delta
y_{t}=f_{t}+\dot{a}f_{\xi}$. As the $t$-dependence in the new
variables is slow,  the term $f_{t}$ can be neglected in this order of the
perturbation scheme. Therefore, we are left with an ordinary
differential equation (the slow time enters as a parameter):
\begin{eqnarray}
(1-u^{2})f_{\xi\xi}+c(2u^{2}+1)f_{\xi} & = &\;
\frac{(c+\dot{a})u}{(1-u^{2})^{1/2}}-\frac{\dot{c}}{c^{2}}q(u)
\nonumber\\&& - \frac{\pi}{2a(1-u^{2})^{1/2}}
 \,.
\label{20}
\end{eqnarray}

Going over from $\xi$ to $u$ and defining $f(\xi)=g(u)$ we obtain
\begin{eqnarray}
(1-u^{2})g_{uu}-3ug_{u} & = &
\frac{(c+\dot{a})}{c^{2}}\frac{1}{u(1-u^{2})^{1/2}}-
\frac{\dot{c}}{c^{4}}\frac{q(u)}{u^{2}} \nonumber\\&&
-\frac{\pi}{2ac^{2}u^{2}(1-u^{2})^{1/2}}
 \,.
\label{21}
\end{eqnarray}
Define  $\Phi(u)=g_{u}(u)$. The general solution of the
homogeneous equation for $\Phi(u)$ is $C\, \Phi_0 (u)$, where
\begin{equation}
\Phi_{0}(u)=(1-u^2)^{-3/2}
 \,,
\label{22}
\end{equation}
and $C$ is an arbitrary constant. Therefore, we look
for the general solution of Eq.
(\ref{21}) in the form
$\Phi(u)=C(u)\Phi_{0}(u)$, where $C(u)$ is an unknown function.
Substituting this into Eq. (\ref{21})
and integrating, we obtain
\begin{equation}
C(u)=\frac{(c+\dot{a})}{c^{2}}\ln\,u-\frac{\dot{c}}{c^{4}}s(u)
+\frac{\pi}{2ac^{2}u}+\alpha(t)
 \,,
\label{24}
\end{equation}
where
\begin{eqnarray}
s(u)&=&\frac{1}{2}\ln^{2}u+\frac{1}{2}\arccos^{2}u -\ln\,u-
\nonumber\\ & &\frac{(1-u^{2})^{1/2}\arccos\,u}{u}
 \,,
\label{24a}
\end{eqnarray}
and $\alpha(t)$ is an arbitrary function of time.  Going back to
$g(u)$, we can write the first order correction to the finger
shape as
\begin{equation}
g(u)=\int_{0}^{u}C(u)\Phi_{0}(u)\,du +\beta(t)
 \,,
\label{25}
\end{equation}
where $\beta(t)$ is another arbitrary function of time. The
boundary conditions for $g(u)$ are $g(0)=0$ and $g(1)=0$. The
first boundary condition gives $\beta(t)=0$, while the second
boundary condition should be used to find an equation for
$\dot{a}$ in the first order. Before doing this, an analysis of
the integrals in Eq. (\ref {25}) should be made. To prevent
divergence of $g(u)$ at $u=1$ we must require
$\alpha(t)=-\pi/(2ca)$. Similarly, to prevent divergence of $g(u)$
at $u=0$ we must require $\dot{c}/c^{2}=-1/a$. The latter equation
is equivalent to Eq. (\ref{Ddot}) obtained in the leading order
from the requirement of area
conservation.

The equation for $\dot{a}$ in the first order of the perturbation
theory is obtained from the boundary condition $g(1)=0$ which
requires calculating the integral in Eq. (\ref {25}) from $u=0$
to $u=1$. This calculation yields
\begin{equation}
\dot{a}=-c+\frac{3\ln2}{a}=-\frac{\pi}{2\Delta}+\frac{3\ln2}{a}
 \,.
\label{26}
\end{equation}
It is now possible to obtain an explicit expression for $g(u)$. An
integration from $0$ to $u$ in Eq. (\ref{25}) gives:
\begin{eqnarray}
g(u)&=&\frac{1}{c^{2}a}
\left[w_{1}(u)-\arcsin\,u\ln\left(\frac{8u}{e} \right) \right.+
\nonumber\\ && \left.
\frac{\pi}{2}\ln\left(\frac{2}{e\left[1+
(1-u^{2})^{1/2}\right]}\right)+w_{2}(u)
 \right]
 \,,
\label{27}
\end{eqnarray}
where
\begin{equation}
w_{1}(u)=\frac{2\,u\,\ln u\,\ln\,
(8u^{1/2}/e)+u\,\arccos^{2}u+\pi(1-u)}{2(1-u^{2})^{1/2}}
 \,,
\label{28}
\end{equation}
and
\begin{equation}
w_{2}(u)=2\int_{0}^{u} \frac{\arcsin\,u}{u}\,du
 \,,
\label{w2}
\end{equation}
In the absence of a conservation law, we would have to go to the
second order of the perturbation scheme to obtain
the equation for $\dot{c}$ in the first order. In our area-preserving
perturbation scheme this
equation can
be obtained already in the first order. Notice that the first-order
correction obeys the following conservation law:
\begin{equation}
\frac{d}{dt}\int_{0}^{a(t)}\delta y(x,t)
\,dx=\dot{a}\,[f(\xi=a,t)-f(\xi=0,t)]=0
 \,,
\label{28a}
\end{equation}
where the last equality results from the boundary conditions. This
relation combined with the requirement that the finger area is
preserved
leads to
\begin{equation}
\frac{d}{dt}\int_{0}^{a(t)}Y(x,t)\,dx=0
 \,.
\label{29}
\end{equation}
It should be stressed that functions $a(t)$ and $\Delta(t)$
entering Eq. (\ref{29}) now include first order corrections. Now,
using Eqs. (\ref{16a}), (\ref{26}) and (\ref{29}) and keeping
terms only up to the first order in $\epsilon$, we obtain
\begin{equation}
\dot{\Delta}=\frac{\pi}{2a}-\frac{\Delta\ln2}{a^{2}}
 \,.
\label{31}
\end{equation}

Equations (\ref{26}) and (\ref{31}) are the equations for
$\dot{a}$ and for $\dot{\Delta}$ in the first order of the
perturbation scheme. This
set of equations is autonomous and therefore integrable.
In the framework of the
perturbation theory it is more consistent to solve the equations
perturbatively by linearization. To do this, we exploit the
conservation law (\ref {29}) which, together with Eq. (\ref{16a}),
allows us to express $\Delta(t)$ as a function of $a(t)$ or vice
versa. The expression obtained is then linearized with respect to
$\Delta/a$ and inserted into Eq. (\ref{26}) (or (\ref {31})). This
yields the following equations:
\begin{equation}
\dot{a}=-\frac{\pi
a}{2a_{0}\Delta_{0}}+\frac{4\ln2}{a}-\frac{a\ln2}{a_{0}^{2}}
 \,,
\label{42a}
\end{equation}
\begin{equation}
\dot{\Delta}=\frac{\pi
\Delta}{2a_{0}\Delta_{0}}+\frac{\Delta\ln2}{a_{0}^{2}}-
\frac{\Delta\ln2}{a^{2}}-\frac{\Delta^{3}\ln2}{a_{0}^{2}\Delta_{0}^{2}}
 \,.
\label{42b}
\end{equation}
We are looking for solutions in the form of
$a(t)=a^{(0)}(t)+a^{(1)}(t)$ and
$\Delta(t)=\Delta^{(0)}(t)+\Delta^{(1)}(t)$, where $a^{(1)}(t)$
and $\Delta^{(1)}(t)$ are first order corrections. Solving the
resulting linear equations, we obtain:
\begin{eqnarray}
a^{(1)}(t)&=&\frac{8\Delta_{0}\ln2}{\pi}\sinh\left(\frac{\pi
t}{2a_{0}\Delta_{0}}\right)-\nonumber\\&&
\frac{t\ln2}{a_{0}}\exp\left(-\frac{\pi
t}{2a_{0}\Delta_{0}}\right)
 \,,
\label{43a}
\end{eqnarray}
\begin{eqnarray}
\Delta^{(1)}(t)&=&-\frac{4\Delta_{0}^{2}\ln2}{\pi a_{0}}
\exp\left(\frac{\pi t}{a_{0}\Delta_{0}}\right)\sinh\left(\frac{\pi
t}{2a_{0}\Delta_{0}}\right)+\nonumber\\&&
\frac{\Delta_{0}t\ln2}{a_{0}^{2}}\exp\left(\frac{\pi
t}{2a_{0}\Delta_{0}}\right)
 \,.
\label{43b}
\end{eqnarray}
We checked that when solving Eqs.(\ref {26}) and (\ref {31})
numerically, the results differ from the approximate analytical
results (\ref{43a}) and (\ref{43b}) by less than $1${\%}. Finally,
requiring $a^{(0)}(t)\gg a^{(1)}(t)$ and $\Delta^{(0)}(t)\gg
\Delta^{(1)}(t)$ and using (\ref{40a}), (\ref{40b}), (\ref{43a}),
and (\ref{43b}), one can see that the linearization procedure is
valid for times obeying the same strong inequality (\ref{41}). Therefore,
the linearization procedure is consistent with the perturbation
theory.

The validity of our perturbative solution is limited by not too
long times,  see inequality  (\ref{41}). On the other hand, one
should wait until irrelevant details of the initial bar shape are
forgotten, and the solution approaches $Y(x,t)$. Typically, it
takes time needed for the bar tip to pass the distance of order of
the characteristic tip size $1/c \sim \Delta$. The tip speed is of
order $1/\Delta$, so the ``waiting time'' of the theory can be
estimated as $\Delta_0^2$. Our perturbative solution represents,
therefore, an intermediate asymptotics, valid for
\begin{equation}
\Delta_{0}^2 \ll t \ll a_{0} \Delta_{0} \, \ln \frac{a_{0}}{\Delta_{0}}\,.
\label{double}
\end{equation}
Finally, we should check the area-preservation criterion (\ref{56}). In
the leading order
we have: $\Lambda(t)\sim a^{(0)}(t)$ and
$\dot{\Lambda}(t)\sim \dot{a}^{(0)}(t)\sim 1/\Delta^{(0)}(t)$.
Using (\ref {40a}) and (\ref {40b}) in (\ref {56}) we obtain:
\begin{equation}
t\ll a_{0}\Delta_{0}\ln(\tilde{f}a_{0})
 \,,
\label{57}
\end{equation}
where $\tilde{f}=(a_0\Delta_{0})/(L_{x}L_{y})$ is the area
fraction occupied by the finger.

\section{numerical solution}

In order to test the predictions of the perturbation theory we
performed numerical simulations with the nonlocal Ginzburg-Landau
equation (\ref{GL}) (with no-flux boundary conditions), taking a
long and narrow rectangular bar as the initial condition. We used
an explicit Euler integration scheme to advance the solution in
time, and second order central differences to discretize the
Laplace operator. A mesh size of $\Delta x=\Delta y=0.5$ was found
sufficient for an accurate resolution of the interface. A time
step of $\Delta t=0.05$ was required for numerical stability. The
bar width $2\Delta_{0} =50$ was chosen to guarantee area
conservation with a good accuracy [see Eq. (\ref{57})]. The
simulation was carried out up to a time $t_{f}=9 \cdot 10^4$
which was long enough to distinguish between an exponential and
linear dynamic behavior for $a(t)$ and $\Delta(t)$. The bar length
$2a_{0}= 8 \cdot 10^3$ was chosen, so that the parameter $\epsilon
(t)=\Delta(t)/a(t)$ was sufficiently small even at the end of
simulation (when it reached $0.1$). The bar was placed in the
center of a rectangular box $8020 \times 220$. Because of symmetry
with respect to the $x$ and $y$ axes only the quadrant $x>0\,,
y>0$ was actually simulated. We checked that the finger area was
conserved with an accuracy better than $1${\%}.

The time dependence of the finger half-length $a(t)$ obtained from
the numerical simulations is shown in Fig. 1. The same graph also
shows the theoretical values of $a(t)$ in the zeroth order
[$a^{(0)}(t)$, Eq. (\ref{40a})], and first order
[$a^{(0)}(t)+a^{(1)}(t)$, Eq. (\ref {43a})] of the perturbation
theory. A good agreement between the numerical and analytical
results is obtained already in the zeroth order. This agreement is
improved further  by the first order correction. For example, at
the final time of the simulation $t_{f}=9 \cdot 10^4$, there is a
$9${\%} difference between the numerical value of $a(t)$ and
$a^{(0)}(t)$, and only a $1.4${\%} difference between the
numerical value of $a(t)$ and $a^{(0)}(t)+a^{(1)}(t)$.

The numerically found
half-width of the finger $\Delta(t)$ is shown in Fig. 2.
Its comparison with the
zero-order values $\Delta^{(0)}(t)$ [Eq. (\ref {40b})]
and with the first-order
values $\Delta^{(0)}(t)+\Delta^{(1)}(t)$ [Eq. (\ref {43b})]
is also shown. Again,
there is a good agreement already in the
zeroth order, and this agreement is further improved by the first
order correction. At the final time of the
simulation, the numerical value of $\Delta(t)$ differs by $6${\%} from
$\Delta^{(0)}(t)$, and by less than $2${\%} from
$\Delta^{(0)}(t)+\Delta^{(1)}(t)$.

\begin{figure}[h]
 \vspace{-0.1in}
\hspace{-1.0cm} \rightline{ \epsfxsize = 8.0cm
\epsffile{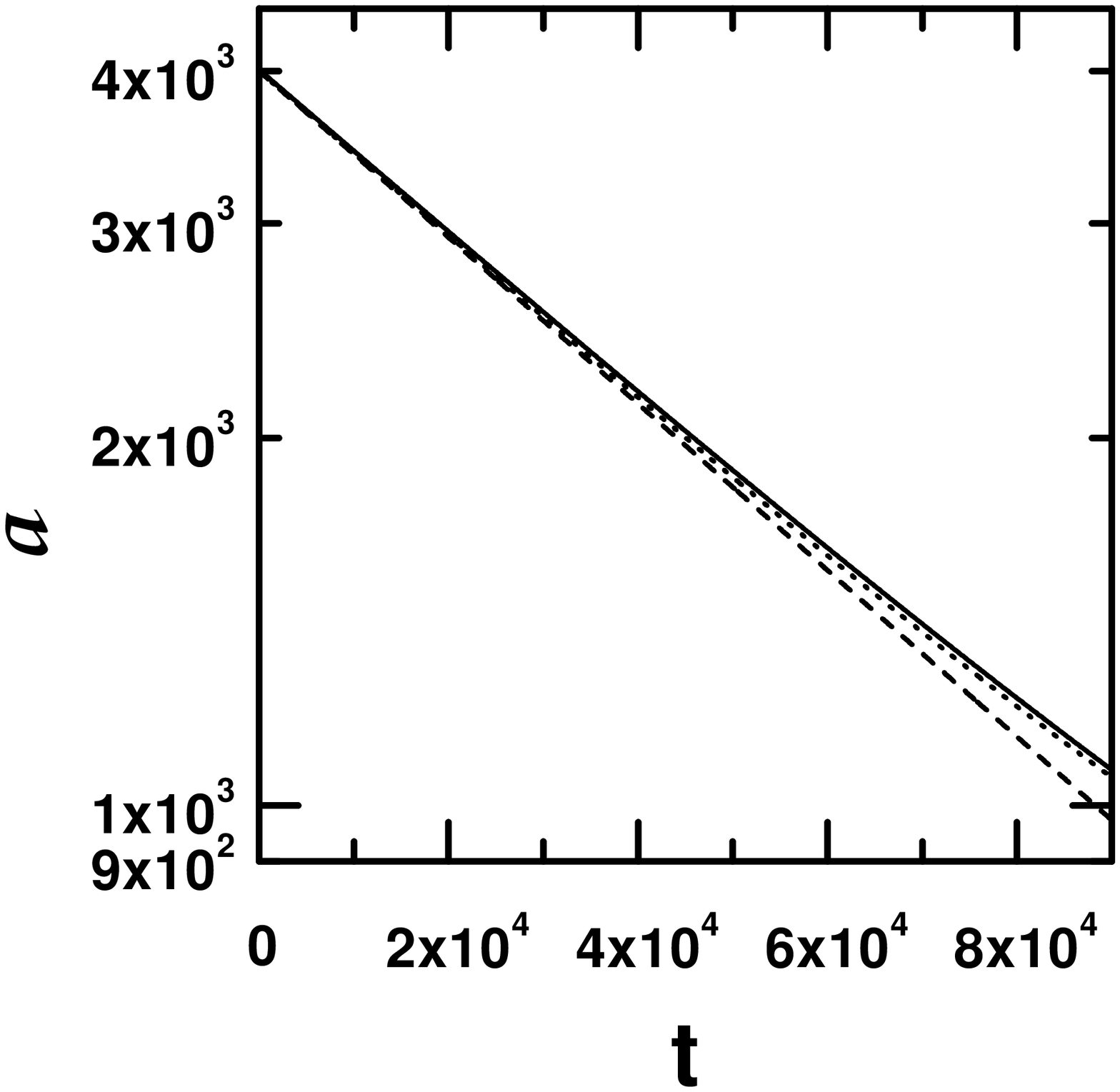}}
 \vspace{-1.2in} \caption{A semi-log plot
of the finger half-length $a$  versus time. The solid line
represents $a(t)$ found numerically. The dashed line shows the
zero-order analytic result $a^{(0)}(t)$. The dotted line is the
first-order analytic result $a^{(0)}(t)+a^{(1)}(t)$.
 \label{fig. 1}}
\end{figure}

\begin{figure}[h]
\hspace{-1.0cm} \rightline{ \epsfxsize = 8.0cm
\epsffile{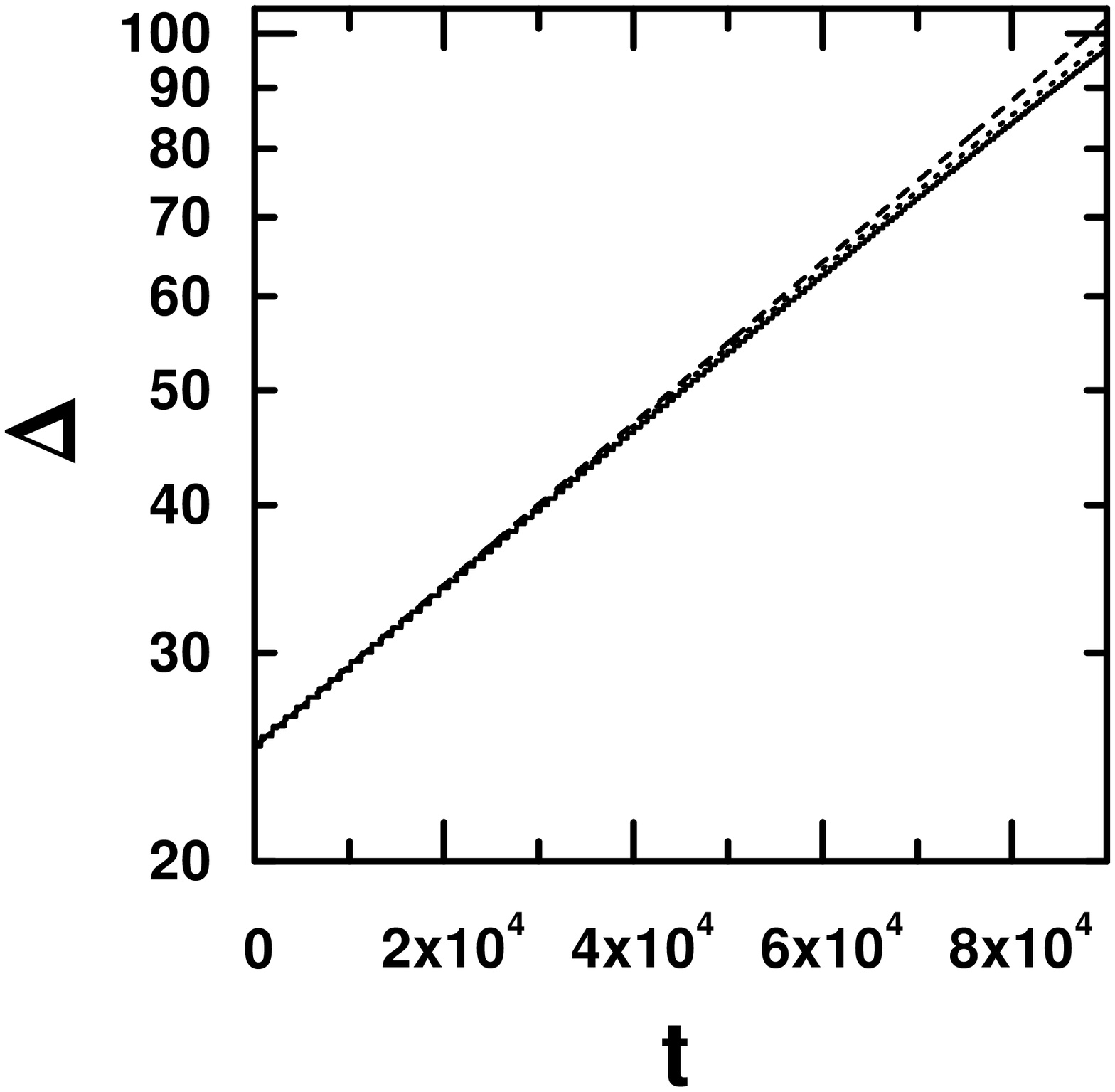}}
 \vspace{-1.2in} \caption{A semi-log plot of
the finger half-width $\Delta$ versus time. The solid line is
$\Delta(t)$ found numerically. The dashed
line shows the zero-order analytic result $\Delta^{(0)}(t)$. The
dotted line is the first-order analytic result
$\Delta^{(0)}(t)+\Delta^{(1)}(t)$.
 \label{fig. 2}}
\end{figure}

Fig. 3 shows the finger shape $y(x,t)$ found numerically, at the
final time of the simulation $t_{f}=9 \cdot 10^4\;$ (notice the
different scales in the $x$ and $y$ axes). Also shown are the
following zero- and first order theoretical results evaluated at
$t=t_{f}$:
\begin{figure}[h]
\vspace{-0.1in} \hspace{-1cm} \rightline{ \epsfxsize = 8.0cm
\epsffile{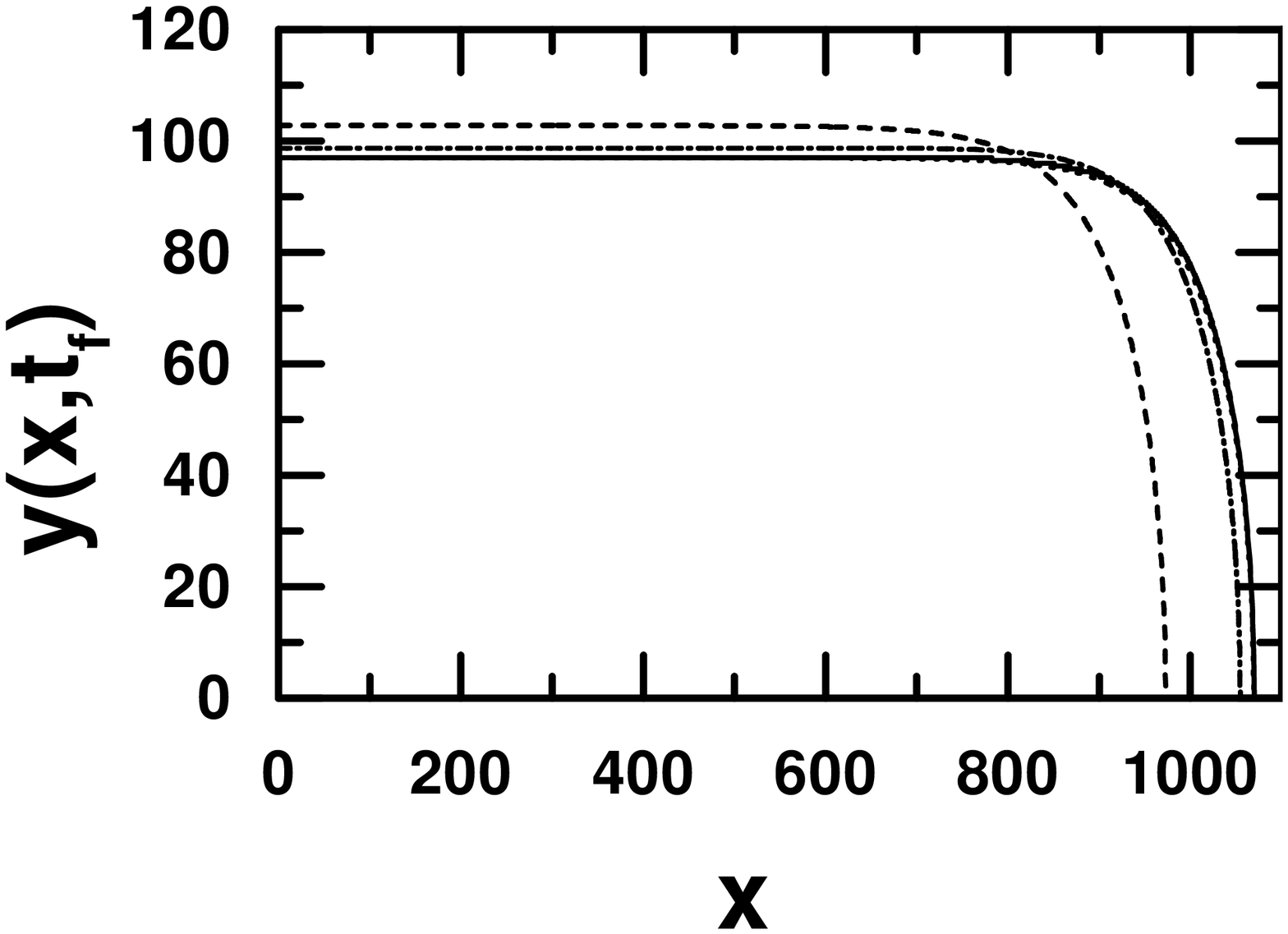}} \vspace{-1.7in} \caption{The finger shape
$y(x,t_{f})$ (solid line). The dashed line is the function
$Y^{(0)}(x,t_f)$ defined by Eq. (\ref{61a}), the dashed-dotted
line is the function (\ref {61b}), and the dotted line is the
function (\ref {61c}). All the functions are evaluated at
$t=t_{f}=9 \cdot 10^4$. The dotted line is indistinguishable from
the solid line. \label{fig. 3}}
\end{figure}
\begin{equation}
Y^{(0)}(x,t)=\frac{1}{c^{(0)}(t)}
\arccos\,[\exp\,(-c^{(0)}(t)\xi^{(0)})]
 \,,
\label{61a}
\end{equation}
where $\xi^{(0)}=a^{(0)}(t)-x$, and
\begin{equation}
y_{s}(x,t)=\frac{1}{c_{s}(t)}
\arccos\,[\exp\,(-c_{s}(t)\xi_{s})]+\delta y(x,t)
 \,,
\label{61b}
\end{equation}
where $\xi_{s}=a_{s}(t)-x$, $a_{s}=a^{(0)}+a^{(1)}$, and
$c_{s}=c^{(0)}+c^{(1)}$. In addition, shown is the function
$Y_{num}(x,t)$ defined by:
\begin{equation}
Y_{num}(x,t)=\frac{1}{c_{num}(t)}\arccos\,[\exp\,(-c_{num}(t)\xi_{num})]
 \,,
\label{61c}
\end{equation}
where $\xi_{num}=a_{num}(t)-x$, and $a_{num}(t)$ and $c_{num}(t)$
are the numerical values of $a(t)$ and $c(t)$, respectively. The
function $Y_{num}(x,t)$ is also evaluated at $t=t_{f}$. The
comparison with the zero-order prediction $Y^{(0)}(x,t_{f})$ shows
a good agreement even at this relatively late time, when $\Delta/a
\simeq 0.1$. The agreement is further improved by the first-order
theory. The comparison with $Y_{num}(x,t_{f})$ shows an excellent
agreement, implying that the subleading term $\delta y(x,t)$ is
very small compared to the leading term $Y(x,t)$. Fig. 4 which
shows the theoretical result for $\delta y(x,t)$ at $t=t_{f}$, calculated by
using Eqs. (\ref {27})-(\ref{w2}), confirms the later conclusion.
We should add that a very good agreement was found
between the theoretical $\delta y(x,t)$ and the numerical $\delta
y(x,t)$, which is obtained by subtracting $Y_{num}(x,t)$ from
the numerical result for $y(x,t)$.

\begin{figure}[h]
 \vspace{-0.1in}
 \hspace{-1cm} \rightline{ \epsfxsize = 8.0cm
\epsffile{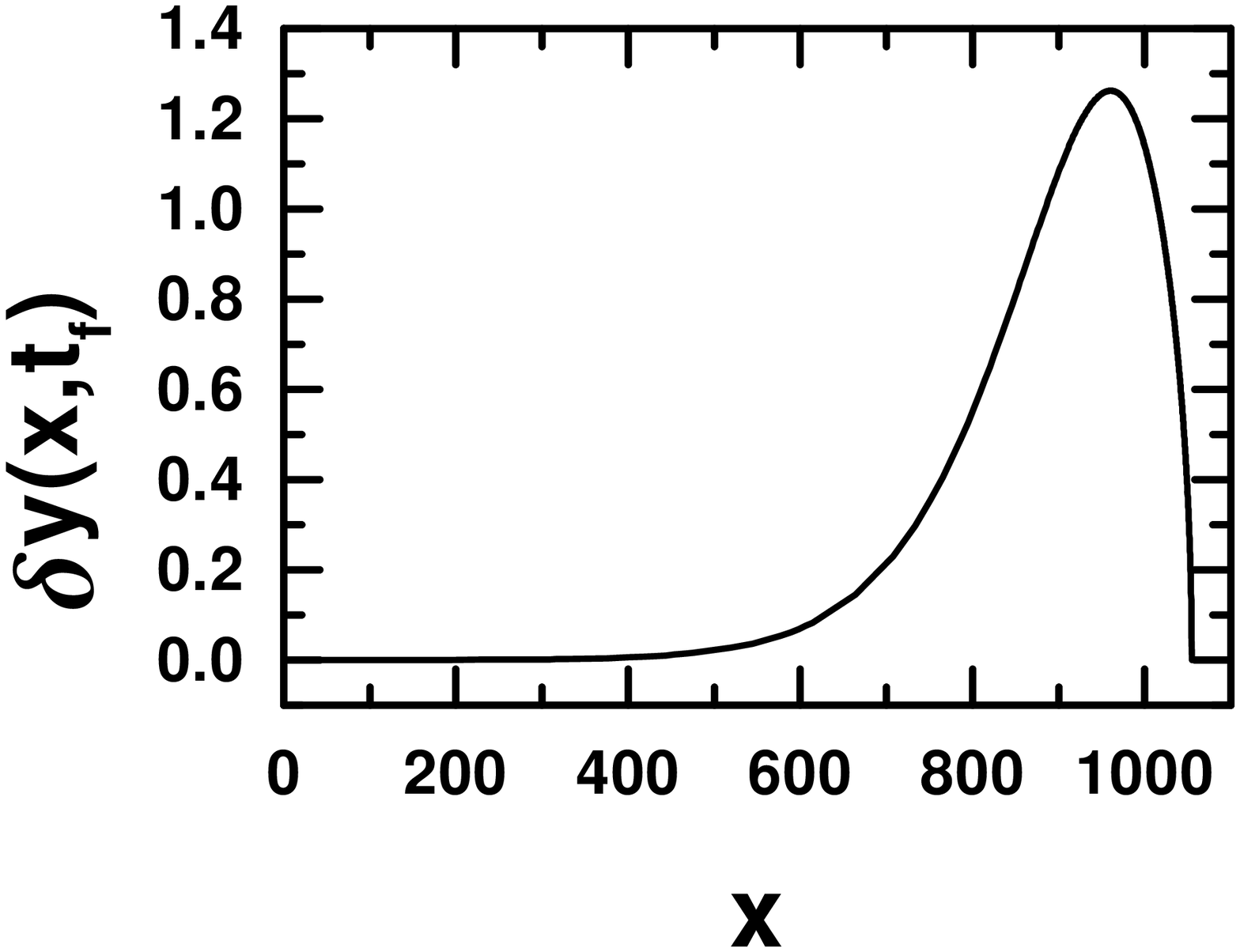}}
 \vspace{-1.7in}
\caption{The subleading term $\delta y(x,t_{f})$ at the final time
of the simulation, as obtained by the perturbation theory [Eqs.
(\ref {27})-(\ref{w2})]. \label{fig. 4}}
\end{figure}

\section{CONCLUSIONS}
We investigated the dynamics of a long slender finger-shaped
domain undergoing globally-conserved (interface-controlled)
coarsening. We worked in the parameter region where these dynamics
can be reduced to area-preserving motion by curvature. The
nonlinear moving boundary problem for the finger shape was solved
by using an asymptotic perturbation theory which employed, as the
zero-order solution, a straightforward modification of the
Mullins solution for a finger retreating by the curvature of its
tip. Both the leading, and the subleading terms of the solution were
calculated analytically and verified by a numerical solution of
the corresponding phase-field equation: the Ginzburg-Landau equation
with a global conservation law.

The coarsening dynamics of the finger look quite differently from
those observed in the case of a locally conserved
(diffusion-controlled) system \cite{CMS2}. The finger preserves
its simple shape, while its characteristic length and width are
changing on the same time scale. Therefore, the global character
of transport (uninhibited by Laplacian screening effects
characteristic of the locally conserved system) manifests itself
already in the simple setting of finger dynamics.

\section*{ACKNOWLEDGEMENTS}
This work was supported in part by the Israel Science Foundation,
administered by the Israel Academy of Sciences and Humanities.

\end{document}